\documentclass[preprint,12pt]{elsarticle}

\usepackage{amsmath}
\usepackage{float}
\usepackage{amssymb}
\usepackage{bm}
\usepackage{graphicx}
\usepackage{booktabs}
\usepackage{multirow}
\usepackage{array}
\usepackage{lineno}
\usepackage{xcolor}
\usepackage{hyperref}
\usepackage{tabularx}
\usepackage{booktabs}
\usepackage{longtable}
\usepackage{pdflscape}
\usepackage{array}
\usepackage{xltabular}
\newcolumntype{Y}{>{\raggedright\arraybackslash}X}

\biboptions{sort&compress}

\graphicspath{{./}}

\newcommand{\LCDM}{\ensuremath{\Lambda\mathrm{CDM}}}
\newcommand{\wCDM}{\ensuremath{w\mathrm{CDM}}}
\newcommand{\FLCDM}{\ensuremath{F\Lambda\mathrm{CDM}}}

\journal{Physics of the Dark Universe}

\begin{document}

\begin{frontmatter}

\title{Constraints on Finsler cosmology from observational data}

\author[aff1]{Bohao Wang}

\author[aff2]{Xiaofeng Yang\corref{cor1}}
\ead{xfyang@henu.edu.cn}

\author[aff2]{Kang Jiao}
\author[aff2,aff3]{Yunliang Ren}
\author[aff2,aff1]{Yangjun Shi}
\author[aff2,aff1]{Jianfeng Meng}
\author[aff2,aff1]{Jingze Li}
\author[aff1]{Xiongwei Liu}

\cortext[cor1]{Corresponding author}

\address[aff1]{School of Physics and Astronomy, China West Normal University,
Nanchong 637002, Sichuan, China}
\address[aff2]{School of Physics and Electronics, Henan University,
Kaifeng 475004, Henan, China}
\address[aff3]{School of Physics and Technology, Xinjiang University,
Urumqi 830046, Xinjiang, China}

\begin{abstract}

We investigate current constraints and conditional detectability of Finslerian \LCDM{} (\FLCDM{}) cosmology using supernova, cosmic chronometer (CC), and baryon acoustic oscillation (BAO) data. 
We adopt PantheonPlus SH0ES, CC and BAO data with a free acoustic scale.
Supplementary tests with DES supernovae and PantheonPlus without SH0ES absolute calibration verify result robustness. 
We compare \FLCDM{} model against {\LCDM} and {\wCDM} and find that \FLCDM{} improves the fit compared to \LCDM{}, but the degree of fit is not superior to that of the $w$CDM model with the same number of parameters.
We find that the background variations introduced by \FLCDM{} can be absorbed to some extent by \LCDM{} through adjusting its own parameters.
We then use parametric bootstrap calibration and Monte Carlo power estimation to assess
the detection power of residual background deviation that \LCDM{} cannot absorb.
We find no statistically significant evidence for a residual background deviation in the main data configuration.
For an injection background fixed at the maximum likelihood estimate, the conditional detection power is relatively low.
We examine the sensitivity of the detection power of this signal to the injected background.
We find that the detection power of this deviation is significantly dependent on the injection context.

\end{abstract}


\end{frontmatter}



\section{Introduction}
\label{sec:introduction}
The standard \LCDM{} cosmological model describes the accelerated expansion of the universe with a small number of parameters and serves as the benchmark model of modern cosmology.
Within this framework, the cosmological principle gives rise to an FLRW background that is homogeneous and isotropic on large scales, while the late-time accelerated expansion is described by a cosmological constant.
Although the reliability of \LCDM{} has been tested by most observations \citep{Planck2018VI,DESI2024VI,Spergel2003,Amanullah2010,Reid2010,Komatsu2011,
Suzuki2012,Hinshaw2013,Betoule2014,Aubourg2015,Scolnic2018,
Aghanim2020Overview,Aghanim2020Likelihoods}, 
the physical nature of dark energy, 
the cosmological constant problem \citep{Weinberg1989}, 
and parameter differences between certain datasets continue to motivate the search for and testing of more general cosmological models.
These models generally fall into two categories.
Dark energy models, which introduce an additional field or fluid,
and extended theories of gravity, which modify Einstein's field equations.

Finsler geometry generalizes Riemannian geometry by relaxing
the quadratic restriction on the line element
\citep{BaoChernShen2000}.
It has been explored as a geometric framework for extending
general relativity
\citep{Takano1974,TavakolVanDenBergh1985,Asanov1985}.
Beil applied Finsler gauge transformations to general relativity
\citep{Beil1992},
while Asanov developed gravitational field
equations based on Finsler geometry
\citep{Asanov1983,Asanov1985}.
Finsler geometry naturally embeds symmetry breaking into the spacetime metric, 
allowing it to describe the asymmetry of spatial orientation without violating the Big Bang framework, 
and it has been widely applied to cosmology\citep{PfeiferWohlfarth2012,ChangLi2008,ChangLi2009,LiChang2010,LiChang2014,WangMeng2018}.

This work uses the separately analyzed PantheonPlus SH0ES and DES-Dovekie supernova compilations,
selected CC measurements and DESI DR2 BAO data 
to test the Finsler-Berwald cosmological model.
We employ reduced background parameterization, 
compare the \FLCDM{}, \LCDM{}, and \wCDM{}.
We find that the \FLCDM{} model has redundant original parameters
and contains a subspace equivalent to \LCDM{}.
We employ the MCMC method to explore the posterior distribution of effective parameters
and determine the range of cosmic background evolution allowed by the observational data.
We then evaluate the portion of the signal that \LCDM{} cannot fully absorb by readjusting its parameters (hereinafter referred to as residual background deviation).
We use parametric bootstrap calibration and Monte Carlo power estimation,
with the maximum likelihood estimate (MLE) as the injection background,
to quantify the detection power under current observational conditions.
We further examine the sensitivity of this power to the choice
of injection background.

This paper is organized as follows.
Section \ref{sec:models} introduces the cosmological models and observational data studied.
Section \ref{sec:methodology} presents likelihood, parameter sampling, model comparison,
calibration and estimation.
Section \ref{sec:results} reports the constraints,
results and sensitivity of the detection power to the background.
Section \ref{sec:conclusion} gives the main conclusions,
outlook and limitations.

\section{Cosmological models and observational data}
\label{sec:models}

\subsection{Finslerian model}
\label{subsec:reference-model}

We use the flat \FLCDM{} model in a Finsler-Berwald spacetime derived by Wang and Meng based on previous research\citep{ChangLi2009,LiChang2010,LiChang2014,WangMeng2017,WangMeng2018}.
The dimensionless expansion rate of this model can be written as
\begin{equation}
E^2(z)
=
\mathcal C_m \Omega_m (1+z)^{\nu_m}
+
\mathcal C_{\rm de}\Omega_{\rm de}
(1+z)^{\nu_{\rm de}},
\label{eq:flcdm-original}
\end{equation}
where
\begin{align}
\mathcal C_m
&=
\frac{9\alpha+\beta+12}
{4(1+\alpha)(3+\beta)},
&
\nu_m
&=
\frac{6(3\alpha+\beta+6)}
{9\alpha+\beta+12},
\label{eq:flcdm-matter-coeff}
\\
\mathcal C_{\rm de}
&=
\frac{9\alpha-\beta+6}
{2(1+\alpha)(3+\beta)},
&
\nu_{\rm de}
&=
\frac{6(3\alpha-\beta)}
{9\alpha-\beta+6}.
\label{eq:flcdm-de-coeff}
\end{align}
Here, $\Omega_m$ and $\Omega_{\rm de}$ are density parameters of matter and dark energy, respectively.
$\alpha$ and $\beta$ are dimensionless Finsler correction parameters.

From the normalization condition $E(0)=1$, we obtain
\begin{equation}
\mathcal C_m\Omega_m
+
\mathcal C_{\rm de}\Omega_{\rm de}
=1,
\end{equation}
which gives
\begin{equation}
\Omega_{\rm de}
=
\frac{1-\mathcal C_m\Omega_m}
{\mathcal C_{\rm de}}.
\label{eq:flcdm-normalization}
\end{equation}
When $\alpha=\beta=0$, $\mathcal C_m=\mathcal C_{\rm de}=1$,
$\nu_m=3$ and $\nu_{\rm de}=0$,
the model recovers the standard flat \LCDM{}.

After eliminating $\Omega_{\rm de}$ in equation~\eqref{eq:flcdm-normalization}, we define
\begin{align}
A
&\equiv
\mathcal C_m\Omega_m
=
\frac{(9\alpha+\beta+12)\Omega_m}
{4(1+\alpha)(3+\beta)},
\label{eq:reduced-A}
\\
u
&\equiv
\frac{3\alpha-\beta}{1+\alpha}.
\label{eq:reduced-u}
\end{align}
Substituting these definitions into equation~\eqref{eq:flcdm-original} gives
\begin{equation}
E^2(z)
=
A(1+z)^{\frac{6(6-u)}{12-u}}
+
(1-A)(1+z)^{\frac{6u}{6+u}}.
\label{eq:flcdm-reduced}
\end{equation}
The reduced expression also has exchange symmetry$(A,u)\longrightarrow(1-A,\,6-u)$.
This transformation exchanges the two power laws in equation ~\eqref{eq:flcdm-reduced}
, without altering $E^2(z)$.
We introduce 
\begin{equation}
(A_c,u_c)=
\begin{cases}
(A,u), & u\leq 3,\\
(1-A,\,6-u), & u>3.
\end{cases}
\label{eq:folded-coordinates}
\end{equation}
to eliminate duplicate representations of computation results.

\subsection{Observational catalogs}
\label{subsec:data}

We treat the Finsler-inspired model as an effective background model
and retain the standard FLRW distance relationship for calculations
and we consider the case of a flat universe.
After being standardized based on their light curves, Type Ia supernovae can provide information on the distance-redshift relation.
We applied the selection criterion 
$
S_i=\mathbf1\{z_{{\rm HD},i}>0.01\ \lor\ {\tt IS\_CALIBRATOR}_i=1\},
$ 
to the public PantheonPlus SH0ES records\citep{scolnic2022,brout2022,riess2022}, retaining 1657 of the 1701 records, including all 77 calibrations.
We also use the 1820 distance-modulus measurements published by DES-Dovekie.
This dataset is a reanalysis of the DES five-year supernova sample,
with updated photometric cross-calibration, light curve model training, and related error handling\citep{popovic2026}.
The distance modulus additive zero point of DES-Dovekie is eliminated as a nuisance parameter.
To investigate the effect of absolute distance calibration,
we also constructed a PantheonPlus analysis without SH0ES calibration.
The selection criterion for this analysis was changed to
$S_i^{\rm no\,SH0ES}
=
\mathbf{1}\left[z_{{\rm HD},i}>0.01\right].$

The cosmic chronometer estimates the cosmic expansion rate through the differential ages of passively evolving galaxies.
We selected $30$ points for analysis, including 26 publicly available CC determinations\citep{Simon2005,Stern2010,Moresco2012,Zhang2014,Moresco2015,Moresco2016,Ratsimbazafy2017,Borghi2022,Tomasetti2023},
a low-redshift measurement\cite{wang2026cc} and three DESI DR1 spectral measurements\citep{Loubser2025}.
We constructed the covariance of the 26 determinations following Moresco et al.\cite{Moresco2020}.
For the three points at $z=(0.46,0.67,0.83)$, we included their correlations using
\begin{equation}
C_{ij}^{(3)}=\rho_{ij}s_i s_j,\quad
s_i=\sqrt{\sigma_{{\rm stat},i}^{2}+\sigma_{{\rm sys},i}^{2}},\quad
\rho=\begin{pmatrix}1&0.932&0.830\\0.932&1&0.776\\0.830&0.776&1\end{pmatrix}.
\end{equation}
The statistical and systematic errors are
\begin{align*}
\boldsymbol\sigma_{\rm stat}&=(0.57,6.39,10.07)\mathrm{km\,s^{-1}\,Mpc^{-1}},\\
\boldsymbol\sigma_{\rm sys}&=(12.32,16.64,15.08)\mathrm{km\,s^{-1}\,Mpc^{-1}}.
\end{align*}
The low-redshift point $H(0.12)=71.33\pm4.20\,\mathrm{km\,s^{-1}\,Mpc^{-1}}$ 
is treated as an independent data.

We used $13$ BAO measurements and complete covariance matrices from DESI DR2\citep{DESI2025DR2I,DESI2025DR2II}, incorporating the sound horizon $r_d$ as a free parameter.

\section{Methodology}
\label{sec:methodology}

MCMC samples the posterior distribution of model parameters by constructing a Markov chain whose stationary distribution is the target posterior.
We used the \texttt{emcee} ensemble sampler to explore parameter posteriors\citep{ForemanMackey2013}
and assessed convergence using the integrated autocorrelation time\citep{Vehtari2021}.
Its commonly used stretch update is based on the Goodman-Weare sampling method,
suitable for handling parameter distributions with strong linear correlation\citep{GoodmanWeare2010}.

The main MCMC configuration in this work is
PantheonPlus SH0ES+CC+BAO,
with the BAO acoustic scale $r_d$ treated as a free parameter.
DES-Dovekie+CC+BAO 
and
PantheonPlus+CC+BAO without SH0ES absolute calibration
are used as sensitivity analyses of data selection and absolute calibration.
We adopted the following uniform priors on the original parameters to explore the posterior structure over a broad parameter range
\begin{align}
H_0
&\sim
\mathcal U
\left(
20,100
\right)
\ {\rm km\,s^{-1}\,Mpc^{-1}},
\\
\Omega_m
&\sim
\mathcal U(0.01,0.99),
\\
\alpha,\beta
&\sim
\mathcal U(-3,3),
\\
r_d
&\sim
\mathcal U(100,180)\ {\rm Mpc}.
\end{align}
For configurations including SH0ES calibration, we adopted
\begin{equation}
M\sim\mathcal U(-20.5,-18).
\end{equation}

The uniform prior is limited to parameter combinations that satisfy the following requirements, $\Omega_{\rm de}>0$, $E^2(z)>0$ (where $0\leq z\leq2.33$), and numerically well-defined background predictions
(invalid parameter values are assigned a logarithmic posterior probability of $-\infty$ by the sampler and rejected during the sampling process).

We employ parametric bootstrap calibration and Monte Carlo power estimation to evaluate the recognition capability of
 the measurement data for residual background deviation signals\citep{Protassov2002,Kashyap2010,Gudicha2016}.
We generated and analyzed two types of mock datasets using the MLE obtained by fitting
\LCDM{} and \FLCDM{} to the real data
as the fixed generating parameters
to
obtain the detection power of the residual background deviation in that injection\cite{DavisonHinkley1997}.
For block $b$ in the $s$-th mock dataset, the simulated data is
\begin{equation}
\mathbf d_b^{(s)}=\mathbf m_b\!\left(\widehat{\boldsymbol{\theta}}_{\mathrm{gen}}\right)+\boldsymbol{\epsilon}_b^{(s)},
\end{equation}
where 
$\mathbf m_b\!\left(\widehat{\boldsymbol{\theta}}_{\mathrm{gen}}\right)$
is the prediction of the generating model evaluated at
\(\widehat{\boldsymbol\theta}_{\rm gen}\).
The noise follows 
$\boldsymbol{\epsilon}_b^{(s)}\sim\mathcal N\!\left(\mathbf 0,C_b\right).$
Here, $(C_b)$
is the covariance matrix corresponding to the data block.

We define the statistical measure of fitting improvement of \FLCDM{} relative to \LCDM{} as
\begin{equation}
q\equiv\max\left\{0,\,\chi^2_{\min,\Lambda}-\chi^2_{\min,F}\right\}.
\end{equation}
Here, $\Lambda$ and $F$ denote \LCDM{} and \FLCDM{}, respectively.
We generated $N_{\rm null}$ null hypothesis mock datasets using the MLE solution of the \LCDM{} fitted to the real data as the generating parameters, and refitted the \LCDM{} and \FLCDM{} on each mock dataset.
The empirical 95th percentile, denoted as $q_{0.95}$, is defined by the distribution of the test statistic $q$ under the null hypothesis, and is set as the significance threshold for determining residual background deviation.
To quantitatively describe the probability of the observed improvement in fitting in real data occurring under a standard universe,
we let the number of the null mock datasets satisfying 
$q^{(s)}\ge q_{\rm obs}$
be
\begin{equation}
r=\sum_{s=1}^{N_{\rm null}}I\!\left(q^{(s)}\ge q_{\rm obs}\right),
\end{equation}
where $I$ is an indicator function, 
and $q_{\rm obs}$ represents the fitting improvement statistic obtained from real observational data:
\begin{equation}
q_{\rm obs}
=
\chi^2_{\min,\Lambda{\rm CDM}}
-
\chi^2_{\min,{\rm F}\Lambda{\rm CDM}}.
\end{equation}
Here, 
$q^{(s)}$ is the statistic corresponding to the $s$-th mock sample under the null hypothesis.
We computed the empirical \(p\)-value using the plus-one correction\cite{PhipsonSmyth2010}:
\begin{equation}
p=\frac{r+1}{N_{\rm null}+1}.
\end{equation}

We then generated $N_{\rm inj}$ mock datasets using the MLE solution obtained from the observational data in the \FLCDM{} model as a fixed injection point.
We fitted \LCDM{} and \FLCDM{} to each injected mock dataset to obtain the injection statistic $q_{\rm inj}^{(s)}$.
We estimated the conditional detection power as follows:
\begin{equation}
Power=\frac{1}{N_{\rm inj}}\sum_{s=1}^{N_{\rm inj}}I\!\left(q_{\rm inj}^{(s)}\ge q_{0.95}\right).
\end{equation}

We used the following methods to examine the degree of dependence of detection power on the background.
We selected the intersection points that satisfy $\Delta\chi^2_{\rm prof}=1$ and $4$ on both sides of the MLE point of the likelihood function.
For these five background points, we optimized $A$, $H_0$, $r_d$, and supernova zero-point parameters under maximum likelihood with $u$ fixed, and evaluated their detection power for residual background deviations.

\section{Results}
\label{sec:results}

\subsection{Constraints and comparisons of models}
\label{subsec:posterior_comparison}
We report the posterior distribution of the parameters
and compare the information criterion results of each model.

\begin{figure}[H]
	\centering
	\includegraphics[width=\linewidth]{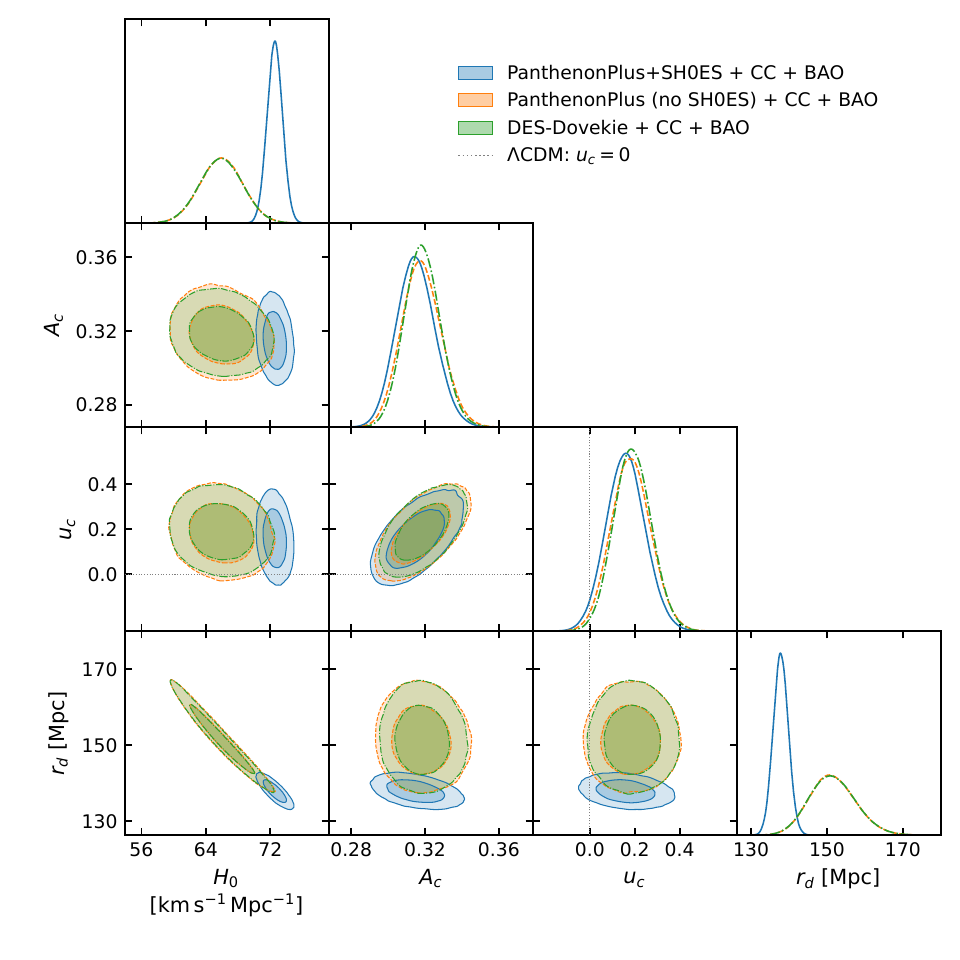}
	\caption{Constraints of \FLCDM{} (68\% and 95\% credible regions).The figure also marks the equivalent background of \LCDM{} with $u_c=0$}
    \label{fig:main}
\end{figure}

The constrained results are presented in Figure~\ref{fig:main}.
We find a positive correlation between $A_c$ and $u_c$,
indicating that compensating changes in the two parameters can preserve similar background predictions.
PantheonPlus SH0ES corresponds to a lower and more concentrated $r_d$.
DES-Dovekie provides median parameters and marginal posterior widths that are similar to those of the PantheonPlus without SH0ES combination.

\begin{figure}[H]
	\centering
	\includegraphics[width=\linewidth]{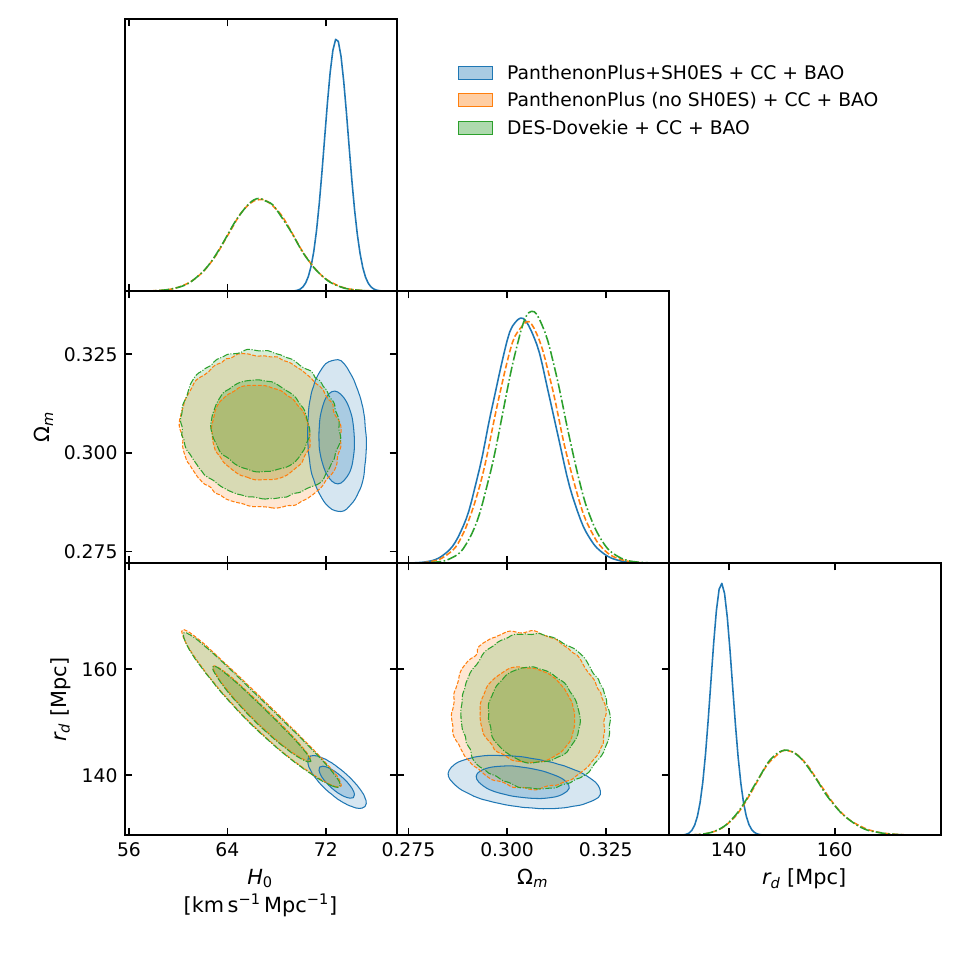}
    \caption{
    Constraints of \LCDM{} (68\% and 95\% credible regions).
    }
    \label{lcdm}
\end{figure}

\begin{figure}[H]
	\centering
	\includegraphics[width=\linewidth]{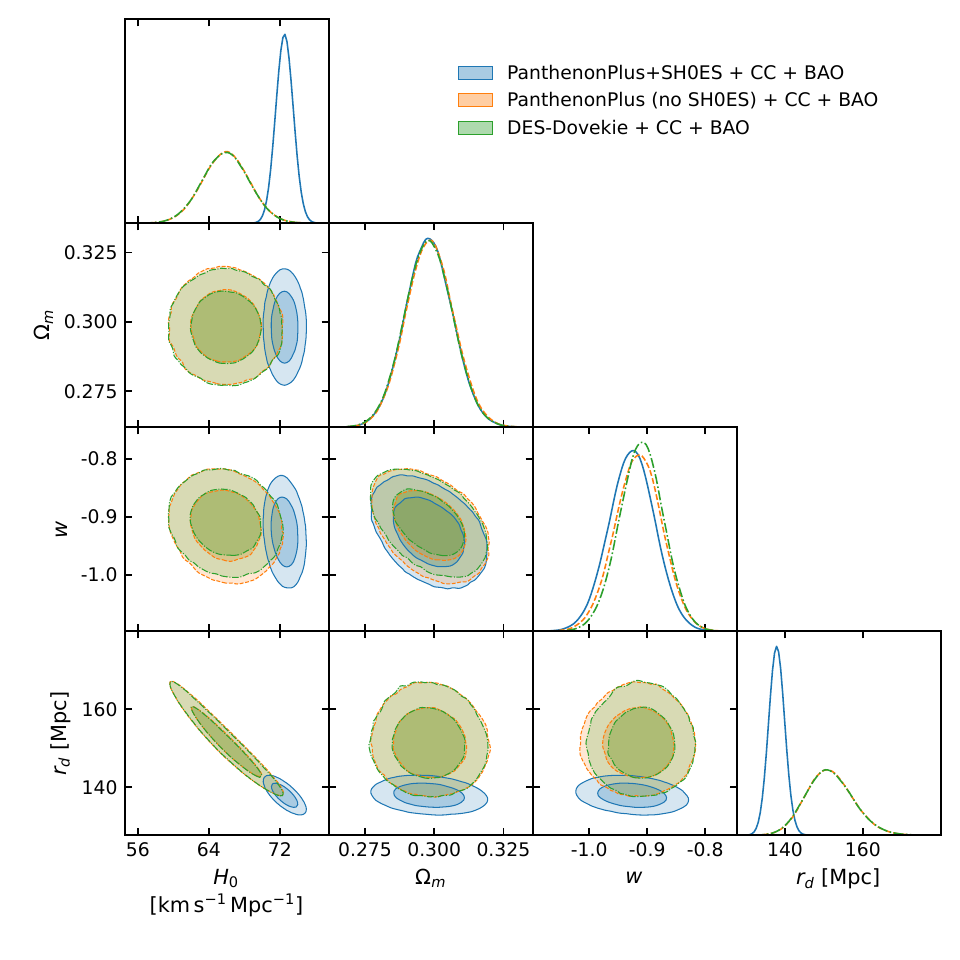}
    \caption{
    Constraints of \wCDM{} (68\% and 95\% credible regions).
    }
    \label{wcdm}
\end{figure}

For comparison,
the parameter constraints obtained from \LCDM{} and \wCDM{} models using the above three data configurations are shown in Figure~\ref{lcdm} and Figure~\ref{wcdm}.
For $\Lambda$CDM, the three data configurations yield similar
constraints on $\Omega_m$, 
with the main differences appearing
in $H_0$ and $r_d$.
PantheonPlus without the SH0ES calibration and DES-Dovekie provide highly overlapping constraints.
In the \wCDM{} model, the free parameter $w$ exhibits a significant negative correlation with $\Omega_m$.
After incorporating this degree of freedom,
the differences in $H_0$--$r_d$ under different calibrations still exist,
and PantheonPlus without SH0ES and DES-Dovekie configurations
still provide a similar joint posterior.

By comparing the parameter constraints we find that
the posterior parameters obtained from the configuration with SH0ES calibration in \FLCDM{} are similar to the results that we obtained in the \LCDM{} and \wCDM{} models.
In the PantheonPlus without SH0ES and DES-Dovekie,
the posterior constraints given by \FLCDM{} are also quite similar.
We also find this phenomenon in  \LCDM{} and \wCDM{} models.
By contrast,
the posterior intervals for $A_c$ and $u_c$ from the three configurations overlap significantly.
The influence of data selection on the scale parameter is more pronounced in \FLCDM{},
while the background shape parameter remains relatively stable.

\begin{table}[tbp]
\centering
\caption{Comparison of posterior constraints and information criteria for models under three sets of data configurations.
The three models have 4, 5, and 5 parameters, respectively.}
\label{tab:posterior_ic_comparison}
\footnotesize
\setlength{\tabcolsep}{2pt}
\renewcommand{\arraystretch}{1.2}
\begin{tabular}{@{}lccccccrr@{}}
\toprule
model & $H_0$ & $\Omega_m$ & $A_c$ & $u_c$ & $w$ & $r_d$ & $\Delta\mathrm{AIC}$ & $\Delta\mathrm{BIC}$ \\
\midrule
\multicolumn{9}{@{}l}{\textbf{PantheonPlus SH0ES + CC + BAO}} \\[2pt]
\LCDM{} & $72.87^{+0.93}_{-0.92}$ & $0.304^{+0.008}_{-0.008}$ & -- & -- & -- & $138.65^{+1.93}_{-1.90}$ & $0.000$ & $0.000$ \\
\FLCDM{} & $72.56^{+0.94}_{-0.94}$ & -- & $0.315^{+0.010}_{-0.010}$ & $0.159^{+0.086}_{-0.084}$ & -- & $137.81^{+1.97}_{-1.94}$ & $-1.237$ & $4.201$ \\
\wCDM{} & $72.54^{+0.94}_{-0.94}$ & $0.298^{+0.009}_{-0.008}$ & -- & -- & $-0.926^{+0.039}_{-0.039}$ & $137.87^{+1.95}_{-1.94}$ & $-1.464$ & $3.975$ \\
\midrule
\multicolumn{9}{@{}l}{\textbf{PantheonPlus without SH0ES+ CC + BAO}} \\[2pt]
\LCDM{} & $66.68^{+2.61}_{-2.62}$ & $0.305^{+0.008}_{-0.008}$ & -- & -- & -- & $151.38^{+6.19}_{-5.72}$ & $0.000$ & $0.000$ \\
\FLCDM{} & $65.93^{+2.60}_{-2.61}$ & -- & $0.318^{+0.011}_{-0.010}$ & $0.182^{+0.088}_{-0.086}$ & -- & $151.27^{+6.20}_{-5.67}$ & $-2.135$ & $3.263$ \\
\wCDM{} & $65.93^{+2.59}_{-2.60}$ & $0.298^{+0.009}_{-0.009}$ & -- & -- & $-0.915^{+0.039}_{-0.040}$ & $151.28^{+6.16}_{-5.66}$ & $-2.435$ & $2.963$ \\
\midrule
\multicolumn{9}{@{}l}{\textbf{DES-Dovekie + CC + BAO}} \\[2pt]
\LCDM{} & $66.65^{+2.62}_{-2.60}$ & $0.307^{+0.008}_{-0.007}$ & -- & -- & -- & $151.25^{+6.14}_{-5.71}$ & $0.000$ & $0.000$ \\
\FLCDM{} & $65.89^{+2.62}_{-2.61}$ & -- & $0.318^{+0.010}_{-0.009}$ & $0.189^{+0.084}_{-0.081}$ & -- & $151.31^{+6.20}_{-5.76}$ & $-3.076$ & $2.454$ \\
\wCDM{} & $65.89^{+2.60}_{-2.61}$ & $0.298^{+0.009}_{-0.009}$ & -- & -- & $-0.910^{+0.037}_{-0.038}$ & $151.26^{+6.20}_{-5.69}$ & $-3.554$ & $1.976$ \\
\bottomrule
\end{tabular}
\medskip

\end{table}

The marginalized posterior medians, $68\%$ credible intervals, and statistical performance of the models are presented in Table~\ref{tab:posterior_ic_comparison}.
Compared to \LCDM{},
the three sets of data exhibit a consistent trend in information criteria:
AIC slightly favors \FLCDM{} and \wCDM{}.
However, BIC favors \LCDM{} in all three data configurations.
The AIC advantage of the Finslerian model is most pronounced in the DES-Dovekie configuration and least pronounced in the PantheonPlus SH0ES.
The noticeable improvement in fitting under the DES-Dovekie configuration may be related to the preference of this dataset combination for additional degrees of freedom in background evolution\cite{popovic2026}, similar trends are also observed in the comparison of $w$CDM.
We find that the information criterion performance of $w $CDM is slightly better than that of \FLCDM.
Due to the same number of parameters in the two models, the difference in AIC and BIC is entirely due to fitting, and the difference is relatively small ($\Delta\mathrm{AIC}\simeq0.23$, $\Delta\mathrm{BIC}\simeq0.23$).
The two information criteria do not provide a consistent preference for either extension over \LCDM{}.

\subsection{Detection of residual background deviation}
\label{subsec:mock_results}

\begin{figure}[H]
	\centering
    
    \includegraphics[width=\linewidth]{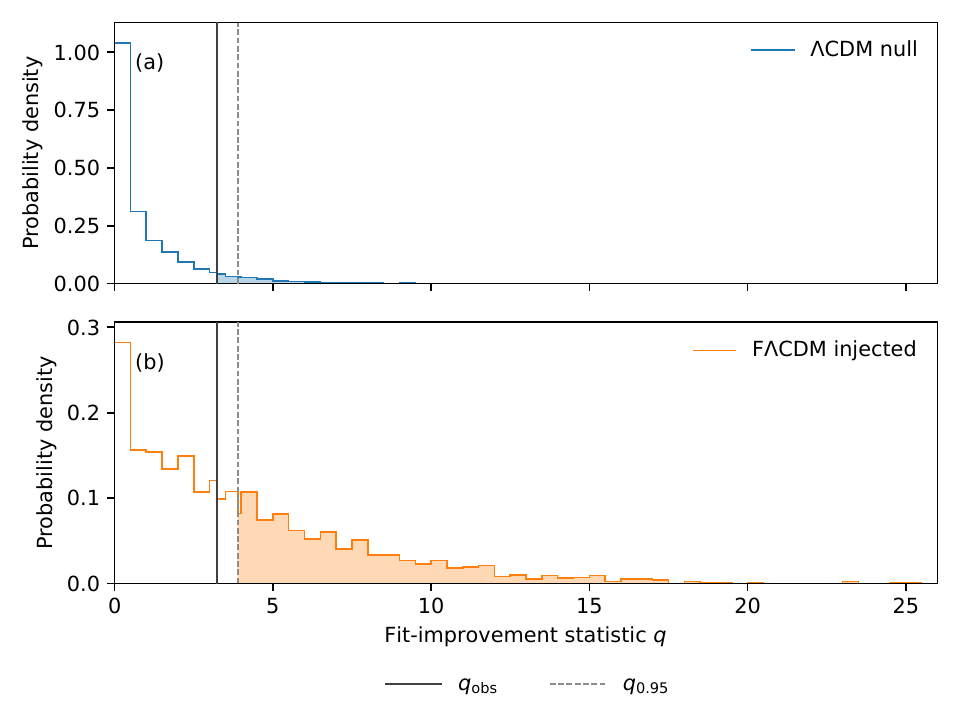}
	\caption{Empirical distributions of the fit-improvement statistic $q$ under the fixed MLE context.
(a) $10,000$ null mock datasets generated from the \LCDM{}. (b) $2,000$ injection mock datasets generated from the \FLCDM{}.
    }
    \label{fig:currentmock} 
\end{figure}

We evaluate the conditional detection power for the residual background deviation under current observational conditions.
The injection point for \FLCDM{} we use is the MLE point,
\begin{align}
(A,u)&=(0.314,0.149),H_0=72.579\,\mathrm{km\,s^{-1}\,Mpc^{-1}},\\
M&=-19.272,\qquad r_d=137.886\,\mathrm{Mpc}.
\end{align}
The mock distributions and numerical results are presented in
Figure~\ref{fig:currentmock} and
Table~\ref{tab:current_detection_power}.
The observed fit improvement lies below the calibrated detection threshold. 
Therefore, 
current data do not provide a statistically significant detection of this deviation.
As a consistency check, we compare the simulated calibration results with the asymptotic distribution of $\chi^2_1$.
At the observed value $q_{\rm obs}=3.24$, the asymptotic reference gives $p_{\rm asym}=0.072$, which is consistent with the simulated value $p_{\rm local}=0.074$ within the Monte Carlo sampling error range.

In the context of MLE injection,
the conditional detection power is only $41.25\%$,
indicating that repeated observations under the assumed background would often fail to reach the detection threshold.
The significant detection in current observational data does not directly imply that the deviation does not exist.

\begin{table}[H]
\centering\small
\setlength{\tabcolsep}{3pt}
\renewcommand{\arraystretch}{1.15}
\caption{Numerical results of the mock simulations
}
\label{tab:current_detection_power}
\begin{tabular}{@{}lr@{}}
\toprule
statistic & results \\
\midrule
$N_{\rm null}$ & $10000$ \\
$N_{\rm inj}$ & $2000$ \\
$q_{\rm obs}$ & $3.24$ \\
$q_{0.95}$ & $3.89$ \\
Null exceedances at $q_{\rm obs}$ & $734/10000$ \\
$p_{\rm local}$ & $0.074$ \\
detection count & $825/2000$ \\
power & $41.25\%$ \\
95\% CI(\%) & $[39.08\%,\,43.44\%]$ \\
\bottomrule
\end{tabular}
\par\smallskip{\footnotesize }
\end{table}

\begin{figure}[H]
	\centering
    
	\includegraphics[width=\linewidth]{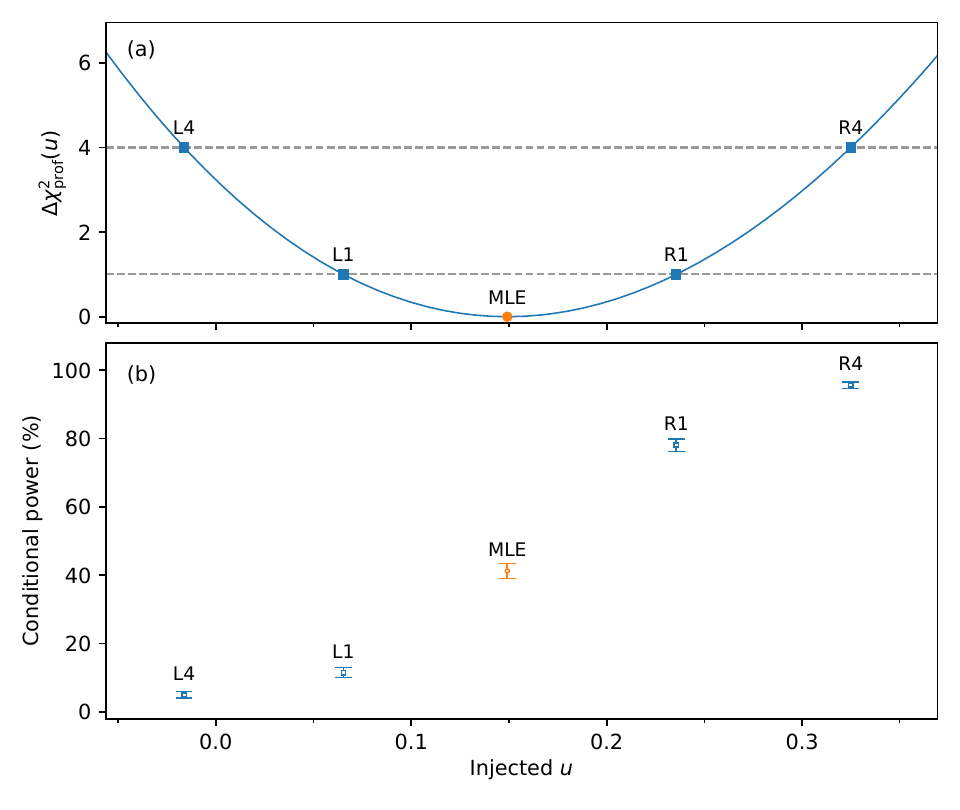}
    \caption{Sensitivity of conditional detection power to the injection background.
(a) Profile likelihood curve of real data, with circular points indicating the MLE and square points indicating the injected backgrounds selected at $\Delta\chi^2_{\rm prof}=1$ and $4$ on both sides. (b) Conditional detection power corresponding to the five backgrounds.
}
\label{fig:sensitivity}
\end{figure}

Since the aforementioned power is only targeted at a fixed MLE injection background,
we next examine its sensitivity to background selection.
The locations of each injection points and corresponding calculation results are shown in Figure~\ref{fig:sensitivity}.
The specific numerical values are listed in Table~\ref{tab:sensitivity}.
With the observational configuration, error covariance, and detection
threshold held fixed, 
the conditional detection power increases
from $4.95\%$ at L4 to $95.70\%$ at R4 along the five selected
injection backgrounds.
This trend of change essentially depends on the position of each injection point in the parameter space relative to the equivalent background of \LCDM{} ($u=0$).
The value of $u\simeq-0.016$ for L4 is already very close to this situation.
The background differences are further compressed after allowing \LCDM{} to readjust its parameters,
and the detection power is approximately $5\%$,
close to the false-positive rate.
From L4 to R4,
the values of $|u|$ at five selected points increase sequentially.
Along the profile path of this investigation,
the background differences that remain unabsorbed after \LCDM{} readjusts its parameters increase,
making it easier for the statistical measure of the simulated data fitting improvement to reach the detection threshold.
Therefore, the detection efficacy gradually increases with the increase of $∣u∣$.

\begin{table}[H]
    \centering
    \small
    \setlength{\tabcolsep}{4pt}
    \renewcommand{\arraystretch}{1.15}
    \caption{
    Numerical results of the injection-background sensitivity tests. The MLE results are retained from the original simulations.
    }
    \label{tab:sensitivity}
    \begin{tabular}{@{}lrrrrrc@{}}
        \toprule
        background
        & $u$
        & $A$
        & $\Delta\chi^2_{\rm prof}$
        & detection count 
        & power
        & 95\% CI\\ \\
        \midrule
        L4
        & $-0.016$ & 0.303 & 4
        & 99/2000 & 4.95\%
        & $[4.04\%,\,5.99\%]$ \\
        L1
        & 0.065 & 0.308 & 1
        & 229/2000 & 11.45\%
        & $[10.09\%,\,12.93\%]$ \\
        MLE
        & 0.149 & 0.314 & 0
        & 825/2000 & 41.25\%
        & $[39.08\%,\,43.44\%]$ \\
        R1
        & 0.236 & 0.320 & 1
        & 1563/2000 & 78.15\%
        & $[76.27\%,\,79.94\%]$ \\
        R4
        & 0.325 & 0.327 & 4
        & 1914/2000 & 95.70\%
        & $[94.72\%,\,96.55\%]$ \\
        \bottomrule
    \end{tabular}
    
\end{table}

\section{Discussion and conclusions}
\label{sec:conclusion}

We combine posterior constraints of effective background parameters, model comparison, and mock calibration
to examine current observational constraints of \FLCDM{}.
In the analysis that uses two separate sets of supernova compilations and combines them with CC and BAO,
the \FLCDM{} model improves the fit.
Both AIC and BIC slightly favor \wCDM{} over \FLCDM{},
which may be due to their different parameter forms. 
\wCDM{} only adjusts the dark energy part through parameter $w$, 
while \FLCDM{} uses $\alpha$ and $\beta$ to describe the global expansion. 
These differences may affect their adaptability to the observational data.
This indicates that the current data do not favor the introduction of more complex Finslerian models.

Further mock analysis indicates that
current observational data do not provide a statistically significant detection of residual background deviation signals.
For an injection background fixed at the MLE point, the conditional detection power is relatively low.
after changing the injection background, the power of five selected backgrounds ranges from $4.95\%$ to $95.70\%$.
This indicates that the detection power of current observations is significantly dependent on the injection.

This study provides a new perspective: for non-standard models with additional background degrees of freedom, posterior constraints on parameters can be further combined with detection power analysis to assess the testability of different parameter backgrounds.
The observed differences retained after parameter re-tuning of the standard model, as well as the dependence of detection efficacy on injection background, should be important considerations in model validation work.
Our sensitivity results in Table~\ref{tab:sensitivity} provide a specific example, 
similar fitting quality does not guarantee similar detection capability.

Our results are based on the standard FLRW distance relations and we neglected cross-covariances between probes.
The accuracy of the current data is limited, and it does not provide restrictions that can significantly distinguish between Finslerian correction and standard background.
We expect future observations with improved precision and broader redshift coverage to further test the model and evaluate its detectability.

\section*{Declaration of competing interest}
The authors declare that they have no known competing financial
interests or personal relationships that could have appeared to influence
the work reported in this paper.

\section*{Acknowledgements}
This work was supported by the National SKA Program of China
(Grants Nos. 2022SKA0110200 and 2022SKA0110203). This work was
also supported by Xiaofeng Yang’s ZHISHAN Distinguished Professor
startup funding of Henan University.

\section*{Data availability}
Data will be made available on request.


\bibliographystyle{elsarticle-num}
\bibliography{cas-refs}

\end{document}